\documentclass[aps,pra,twocolumn,floatfix,showpacs]{revtex4-1}
\usepackage{graphicx,amsfonts}
\usepackage{amsmath,amscd,amsfonts,amssymb,color}
\newtheorem{theorem}{Theorem}

\begin{document}
\title{Extremal extensions of entanglement witnesses and
their connection with UPB}
\author{R. Sengupta}
\email{ritabrata@iisermohali.ac.in}
\affiliation{Indian Institute of Science Education
\& Research Mohali (IISERM), India}
\author{Arvind}
\email{arvind@iisermohali.ac.in}
\affiliation{Indian Institute of Science Education \&
Research Mohali (IISERM), India}
\begin{abstract}
In this paper we describe a new connection between UPB
(unextendable product bases) and P (positive) maps which are
not CP (completely positive).  We show that inner
automorphisms of the set of P maps which are not CP, produce
extremal extensions of these maps that help in
entanglement detection.  By constructing such an extension
of the well-known Choi map, we strengthen its power to
unearth PPT (positive under partial transpose) entangled
states.  We further show that  the class of maps generated
from the Choi map via an inner automorphism naturally detects the
entanglement of states in the orthogonal complement of
certain UPB.
This brings out a hitherto undiscovered connection between
the Choi map and UPB.  We also show that certain  other recently
considered extremal extensions are obtainable by such
extensions of the Choi map.  
\end{abstract} 
\maketitle
\section{Introduction}
\label{introduction}
Quantum entanglement is a  fundamentally new feature that
emerges in the quantum world and its study remains a central
theme in quantum theory.  On the one hand, entanglement  is
responsible for the non-classical correlations leading to
the violation of Bell type inequalities and on the other, it
plays a key role in quantum algorithms giving them a
clear advantage over their classical counterparts\cite{NC}.

In the quantum mechanical description, the physical states
of the system are represented by trace class operators
denoted by $\rho$ on a complex Hilbert space $\mathcal{H}$.
The dimension of this space can be finite or infinite and
for the finite dimensional case, which concern us in this
paper, we have $\mathcal{H}=\mathbb{C}^n$.  If the rank of
$\rho$ is $1$ the state is pure otherwise it is mixed.  The
set of states forms a convex set with pure states being the
extremal points.

For composite systems the Hilbert space is the tensor
product of the Hilbert spaces of the individual systems.
Thus the state space of a bipartite system is given by
$\mathcal{B}(\mathcal{H}_A\otimes\mathcal{H}_B)$.  A
bipartite state $\rho\in
\mathcal{B}(\mathcal{H}_A\otimes\mathcal{H}_B)$ is called a
separable state if and only if it can be written as 
\begin{equation}
\rho=\sum_{j=1}^n p_j
\rho_j^A\otimes\rho_j^B,\quad p_j>0,\quad
\sum_{j=1}^np_j=1.
\label{separable}
\end{equation}
where $\rho_j^A$ and $\rho_j^B$ are states in the systems
$A$ and $B$ respectively.  If a state
$\rho\in\mathcal{B}(\mathcal{H}_A\otimes\mathcal{H}_B)$
cannot be written in the above form, then it is an entangled
state.

The central question in this field is to determine whether a
given arbitrary (pure or mixed) bipartite state $\rho$  is
entangled or separable.  The problem has a simple solution
for the case of pure states.  A pure bipartite state is
separable if and only if the reduced density operator
obtained by tracing over one of the systems is pure. In fact
the entropy of the reduced density operator can be used to
quantify the amount of entanglement. However, for the case
of mixed states such a characterization is not possible and
only partial solutions are available. While there are methods
to uncover entangled states, all of them involve one-way
conditions whose violation indicates entanglement.  The sum
total of such conditions leading to a complete
characterization of states is not available and the solution
to this problem has remained elusive.  A vast body of
literature exists in this field and for a review
see~\cite{RevModPhys.81.865,guth}.

In this process of distinguishing entangled states from
separable ones, the most important mathematical tool is
provided by positive (P) maps which are not completely
positive (CP)~\cite{H1}.  A map
$\varphi:\mathcal{B}(\mathcal{H})\longrightarrow\mathcal{B}(\mathcal{H})$
is said to be positive if it maps the set of positive
operators in $\mathcal{B}(\mathcal{H})$ (denoted by
$\mathcal{B}(\mathcal{H})_+$) to itself. A positive map is
said to be completely positive if the extension
$1_d\otimes\varphi:\mathcal{B}(\mathbb{C}^d\otimes
\mathcal{H})\longrightarrow\mathcal{B}(\mathbb{C}^d\otimes
\mathcal{H})$ is a positive map for all $d \geq 1$.  While
CP maps show a remarkably simple representation due to
Sudarshan, Kraus and Choi~\cite{sud1,kraus,choi1}, P maps
which are not CP  are not easily characterizable. The fact
that all separable states defined by
Equation~(\ref{separable}) remain positive when we apply P
maps which are not CP to one of the systems, helps us to
convert such maps into entanglement witnesses. Therefore,
any state which turns into a non-state under the application
of a P map which is not CP on one of the systems, has to be
entangled. Thus such  maps help us detect bipartite
entangled states.  The transpose operation  was the first  such
P map which is not CP used to detect entanglement.
Using the results of Arveson~\cite{arve69, arve74} and
St{\o}rmer~\cite{sto1, sto2}, Woronowicz~\cite{wor} showed
that in the dimensions $2\otimes2$ and $2\otimes3$ the
transpose map is powerful enough to detect all entangled
states and we need not use any other witnesses. For
composite systems with dimensions larger than $2\otimes 3$,
the
transpose is a useful tool to detect entangled states,
however it does not detect all the entangled states. In
higher dimensions, the states which are negative under
partial transpose (NPT) are entangled while those which
are positive under partial transpose (PPT) can either be separable
or entangled. For the PPT entangled states there has to
exist a P map (not CP) which will convert them into a
non-state and provide a witness for their entanglement.
The Choi map is the first non-trivial example of a P map
which is not CP and which unearths entanglement of PPT
entangled states~\cite{choi2,sto2}.  Only a few more
examples of such maps are available in the literature and the
theory of such maps is far from
complete~\cite{cho1,kye,os,ha1,ha2,rob1,bru1,bru2, hall1,
kossak4, T1,1751-8121-44-21-215305,
1751-8121-45-39-395307,1751-8121-45-41-415305,
PhysRevA.86.034301}.  In the absence of a complete solution,
discovering new families of PPT entangled states, finding
new entanglement witnesses, and understanding their
connections and structure is important.

It has been shown that for a composite system with dimension
more than $2\otimes 3$, one can construct a set of orthogonal
product states spanning a subspace,
such that there is no product state in the orthogonal
complement of this subspace~\cite{B2}. This implies that any state in the
orthogonal complement of this set is an entangled state.
Since these states are by construction PPT, this allows one
to construct families of PPT entangled states~\cite{B2}.
Such families of PPT entangled states should become
non-states when some P map which is not CP acts on one of the
subsystems, thereby revealing their entanglement. Examples
of such maps, although existing in literature, are somewhat
contrived~\cite{T1}.  We show that these families  of PPT
entangled states in the orthogonal complement of UPB have a
connection with the Choi map. This connection is new and
provides insights into the Choi map as well as UPB.

In this work we consider the extremal extension of the
positive maps. We show that this changes their ability to
detect entanglement. We begin with the Choi map which is an
extremal map and construct its extremal extensions using
appropriate automorphisms.  To the best of our knowledge,
the Choi map~\cite{choi2} and its extension~\cite{choi1} are
the only examples of maps which are unital, extremal and
exposed. Our extensions based on automorphisms preserve the
extremality and exposedness and we can always restrict the
extensions to a
sub-class of automorphisms to preserve the unital nature of
the map.  The family of extremal extensions thus generated
are expected to be able to reveal entanglement of new
classes of states.  We then define a one-parameter family of
such extremal extensions and show that for a certain value
of the parameter, the map is able to implicate the
entanglement of states in the orthogonal complement of UPB
arising out of the TILES and PYRAMID construction~\cite{B2,T2}.
This demonstrates that, where the original Choi maps fail to
reveal entanglement of states based on UPB, their extremal
extensions succeed.

The material in this paper is arranged as follows: In
Section~\ref{construction} we describe the construction of
extremal extensions of the P maps which are not CP. We show
how the extensions preserve extremality and how only inner
automorphisms are useful in the context of entanglement
detection. In Section~\ref{choi-upb} we take the example of
the Choi map and construct its extremal extensions using the
method described in Section~\ref{construction}, while
restricting ourselves to $3\otimes 3$ systems. We define a
very interesting family of these extensions where the
quantum operation is restricted to a one-parameter subgroup
of $SU(3)$.  We show that this particular family, for
certain values of the parameters, is able to detect
entanglement of PPT entangled states in orthogonal
complement of UPB for TILES and PYRAMID constructions.  In
Section~\ref{examples} we describe three more examples of
automorphisms demonstrating the usefulness of the
formulation.  Section~\ref{conclusions} contains some
concluding remarks.
\section{Extremal extensions of Positive Maps}
\label{construction}
In this section, starting with a P map (which is not CP) and
a CP  map,  we construct a composite map.  This composite
map turns out to be extremal if the original map is extremal
and under certain conditions has more power to detect
entanglement as compared to the original map.  Consider
$\varphi:\mathcal{B}(\mathcal{H})
\longrightarrow\mathcal{B}(\mathcal{H})$ to be a positive
indecomposable map. 
For any $A\in Gl_n(\mathbb{C})$ we can  define a map
\begin{align}
A:\mathcal{B}(\mathcal{H})&\longrightarrow
\mathcal{B}(\mathcal{H}) \nonumber \\
X&\longmapsto AXA^\dag \quad 
{\rm For}\,\, X\in \mathcal{B}(\mathcal{H})
\end{align}
Clearly, $A$  is a CP map. Note that we are using the same
symobol $A$ for the $Gl_n(\mathbb{C})$ element and the
corresponding map. To make it a valid
quantum operation, we  impose the condition
$AA^\dag\leq I$ where $I$ denotes the
identity element of $\mathcal{B}(\mathcal{H})$.

We can then define the two  automorphisms as the
compositions
\begin{align}
\varphi\circ
A=&\varphi_A \nonumber  \\
A\circ
\varphi=&\varphi^A
\end{align}
The former is called inner automorphism while the
latter is called outer automorphism. The outer
automorphism is not useful for us as it does not
strengthen the entanglement detection capability
of $\varphi$. However as we will see below and in the
next sections, the inner automorphism is
useful. 

It is worth noting that the set of positive maps
is a convex set and can be described by its
`extremal points',  in our case `extremal maps'.
A positive map $h$ is said to be extremal, when
for any decomposition $h=h_1+h_2$, where $h_1$ and
$h_2$ are positive maps, $h_i=\lambda_i h$, where
$\lambda_i\geq0$ and $\lambda_1+\lambda_2=1$.

\begin{theorem}\label{extre1}
For any positive map
$\varphi:\mathcal{B}(\mathcal{H})
\longmapsto\mathcal{B}(\mathcal{H})$,
and for any full rank operator $A$, (such that
$AA^\dag\leq I$) $\varphi_A$ is a positive map.
Moreover, if $\varphi$ is not completely positive and
extremal, so is the map $\varphi_A$.  \end{theorem}

{\bf Proof:}
The map $A:X\longmapsto A X A^\dag$, when $A$ is a
non-singular operator defines an automorphism on
$\mathcal{B}(\mathcal{H})$. If $X$ is Hermitian,
so is $AXA^\dag$ and if $X$ is positive, so is the
image as the map A is completely
positive. Thus the map $A$ is a bijection map from
the set of positive semi-definite operators onto
itself.

Let $\varphi$ be a P but not CP map.  Assume
that $\varphi_A$ is a CP map. Then by Kraus
decomposition, there exists a finite set of
operators $\{V_i\}$ which represents the map and
we can write for any
$X\in\mathcal{B}(\mathcal{H})$ 
\begin{equation}
\varphi_A(X)=\sum_i
V_iXV_i^\dag. 
\end{equation}
Now $\varphi(X)=\varphi_A(A^{-1}X{A^\dag}^{-1})$ since
$A$ is a non-singular operator. We thus have 
\begin{equation}
\varphi(X)=\varphi_A(A^{-1}X{A^\dag}^{-1})=\sum_i
V_iA^{-1}X{A^\dag}^{-1}V_i^\dag .
\end{equation}
implying that $\varphi$ is a CP map. This is a
contradiction.  Hence $\varphi_A$ is a P but not CP
map.

For the second part, let $\varphi$ be extremal
and  let us assume that $\varphi_A$ is not extremal. 
Then there exist positive maps $\varphi^1$ and 
$\varphi^2$ so that
$\varphi_A=\varphi^1+\varphi^2$. 
Using a similar argument as above we
can write
\begin{eqnarray}
\varphi(X)&=& \varphi_A(A^{-1}X{A^\dag}^{-1}) \nonumber \\
	&=& \varphi^1(A^{-1}X{A^\dag}^{-1}) 
+ \varphi^2(A^{-1}X{A^\dag}^{-1}) \nonumber \\
	& = &\varphi_{A^{-1}}^1(X)
+\varphi_{A^{-1}}^2(X).\end{eqnarray}
But the map $\varphi$  is an extremal map.  By definition of
extremality, if
$\varphi=\varphi_1+\varphi_2$, where $\varphi_i$ are positive maps, then
$\varphi_i=\lambda_i\varphi$, where $\lambda_1+\lambda_2=1$. Hence
\begin{equation}
\begin{split}
\varphi_{A^{-1}}^i=\lambda_i\varphi & \Rightarrow
\varphi_{A^{-1}}^i\circ A=\lambda_i\varphi\circ A\\
	& \Rightarrow \varphi^i=\lambda_i\varphi_A.
\end{split}
\end{equation}
Hence $\varphi_A$ is an extremal map.

A special case of interest is when $A$ is unitary which we
denote by $U$. A number of special results are available for
this case.  By the Russo-Dye theorem (see~\cite{bhatia1}) we can
show that  for any unitary operator $U$, 
\begin{equation}
\|\varphi^U\|=\|\varphi_U\|=\|\varphi\|.
\end{equation}
It is ovious that 
if $\varphi$ is unital, so are $\varphi_U$ and $\varphi^U$.

Further, positivity under partial transpose is invariant
under inner unitary automorphism. In other words, for the
transpose map  $T$ and any unitary operator $U$, $(I\otimes
T)\rho\geq0$ implies $(I\otimes T_U)\rho\geq0$ for any state
$\rho$. 

This can be proved as follows:
Let $\rho\in\mathcal{B}(\mathcal{H}\otimes\mathcal{H})$ be a
PPT state. Let us write
$\rho=((\rho_{ij}))$ in the block form, where for each $i$ and $j$,
$\rho_{ij}\in\mathcal{B}(\mathcal{H})$. Then $(I\otimes
T)\rho=((T(\rho_{ij}))=((\rho_{ij}^T))$. Hence 
\begin{eqnarray}
(1\otimes
T_U)\rho &=&((T(U\rho_{ij}U^\dag))\nonumber 
\\
&=&((\overline{U}T(\rho_{ij})\overline{U}^\dag))\nonumber \\ 
&=& 
(I\otimes\overline{U})(1\otimes
T)\rho(I\otimes\overline{U})^\dag.
\end{eqnarray}
Where $U=((u_{ij}))$ and its complex congugate 
$\overline{U}=((\overline{u_{ij}}))$  are
unitary operators.
Since eigenvalues remain invariant
under unitary transformations (local unitary in
our case), the result follows. 
\begin{theorem}
\noindent\begin{enumerate}
\item For any positive map
$\varphi:\mathcal{B}(\mathcal{H})\longrightarrow\mathcal{B}(\mathcal{H})$,
and any unitary operator $U$, the outer automorphism
$\varphi^U$ is a positive map.  \item Any entangled state
$\rho$ detected by $\varphi^U$ is detected by $\varphi$ and
vice versa.  \end{enumerate}
\end{theorem}
{\bf Proof:}
Let $x\in\mathcal{B}(\mathcal{H})$ be any positive
semi-definite Hermitian operator.  Since $\varphi$ is
positive, $\varphi(x)\geq0$. Since the unitary operators do
not change eigenvalues, we have $U\varphi(x)U^\dag\geq0$,
i.e.  $\varphi^U=U(x) \circ \varphi \geq$. Hence $\varphi^U$
is a positive map.

For the second part, notice that the eigenvalues are
invariant under unitary operators. Hence,
\begin{eqnarray}
(1\otimes\varphi)\rho\not\geq0 &\Longleftrightarrow& 
(I\otimes
U)(1\otimes\varphi)\rho(I\otimes
U)^\dag\not\geq0 \nonumber \\
&\Longleftrightarrow&(1\otimes U\varphi
U^\dag)\rho\not\geq0 \nonumber \\
&\Longleftrightarrow&
(1\otimes\varphi^U)\rho\not\geq0.
\end{eqnarray}
This means that for the entanglement detection application,
unitary outer automorphisms are not useful and therefore we
should focus only on the inner automorphism.

In the next section we discuss the power of such
extensions. We will consider PPT entangled states discovered
through UPB construction due to Bennett et. al.~\cite{B1}
and apply one-parameter sub-families of unitary inner
automorphisms to them.
\section{Extensions of Choi map and UPB construction}
\label{choi-upb}
\subsection{The Choi Map}
The first non-trivial example of a P map which is
not CP and can provide a witness for the
entanglement of some PPT entangled states was
discovered by Choi~\cite{choi2}. This map comes in
two variants and they are defined on a
$3$-dim Hilbert space as follows: 
\begin{equation}
\varphi_{C_{1}}:((x_{ij}))\longmapsto\frac{1}{2}\begin{pmatrix}
x_{11}+x_{22} & -x_{12} & -x_{13}\\
-x_{21} &x_{22}+x_{33} & -x_{23}\\
-x_{31} & -x_{32} & x_{33}+x_{11}
\end{pmatrix}
\label{choi1}
\end{equation}
and
\begin{equation}
\varphi_{C_2}:((x_{ij}))\longmapsto\frac{1}{2}\begin{pmatrix}
x_{11}+x_{33} & -x_{12} & -x_{13}\\
-x_{21} &x_{22}+x_{11} & -x_{23}\\
-x_{31} & -x_{32} & x_{33}+x_{22}
\label{choi2}
\end{pmatrix}
\end{equation}
Both these maps as defined in~(\ref{choi1})
and~(\ref{choi2})
are useful in unearthing entanglement of PPT entangled
states and are extremal points in the space of 
maps~\cite{choilam}.  There are only a few examples of
extremal maps and apart from Choi maps, there have been
extensions of Choi maps by Kye~\cite{kye} which were shown to
be extremal by Osaka~\cite{os}.  We are interested in
unitary inner automorphisms of the Choi maps which are
defined as the composition $\varphi_{C_{1,2}} \circ U $ where
$U \in SU(3)$ is a unitary operator. For every $U \in SU(3)$
we have an extremal map generated from the Choi map. For
example, for every one-parameter subgroup of $SU(3)$ we will
have a family of maps which can help us unearth entanglement
of PPT entangled states. 
\subsection{The TILES construction}
The unextendable product basis, the `TILES' construction was proposed by 
Bennett et. al.~\cite{B2}. 
Given a composite system with Hilbert space 
$\mathbb{C}^3\otimes\mathbb{C}^3$, we consider the  
normalized orthogonal states
\begin{eqnarray}
|\psi_0\rangle
&=&\frac{1}{\sqrt{2}}|0
\rangle\left(|0\rangle-|1\rangle\right),\quad
|\psi_2\rangle =
\frac{1}{\sqrt{2}}|2\rangle
\left(|1\rangle-|2\rangle\right),
\nonumber \\
|\psi_1\rangle &=&
\frac{1}{\sqrt{2}}
\left(|0\rangle-|1\rangle\right)
|2\rangle,\quad
|\psi_3\rangle =
\frac{1}{\sqrt{2}}
\left(|1\rangle-|2\rangle\right)|0\rangle, \nonumber \\
|\psi_4\rangle &=&
\frac{1}{3}
\left(|0\rangle+|1\rangle+|2\rangle\right)
\left(|0\rangle+|1\rangle+|2\rangle\right)
\end{eqnarray}
Bennet et. al.  showed that there is no product state in the
orthogonal complement of these states. Therefore,
the state  
\begin{equation}
\rho=\frac{1}{4}\left(I_9-\sum_{i=0}^4|\psi_i\rangle
\langle\psi_i|\right).
\label{upb_entang}
\end{equation}
is entangled. Further, by construction this state is PPT and
therefore we have a PPT entangled state.  We can apply the
maps $I\otimes \varphi_{C_{1,2}}$ to the state and it turns out
that the state remains positive and does not reveal
its entanglement.
\begin{figure}[h]
\includegraphics[width=7cm]{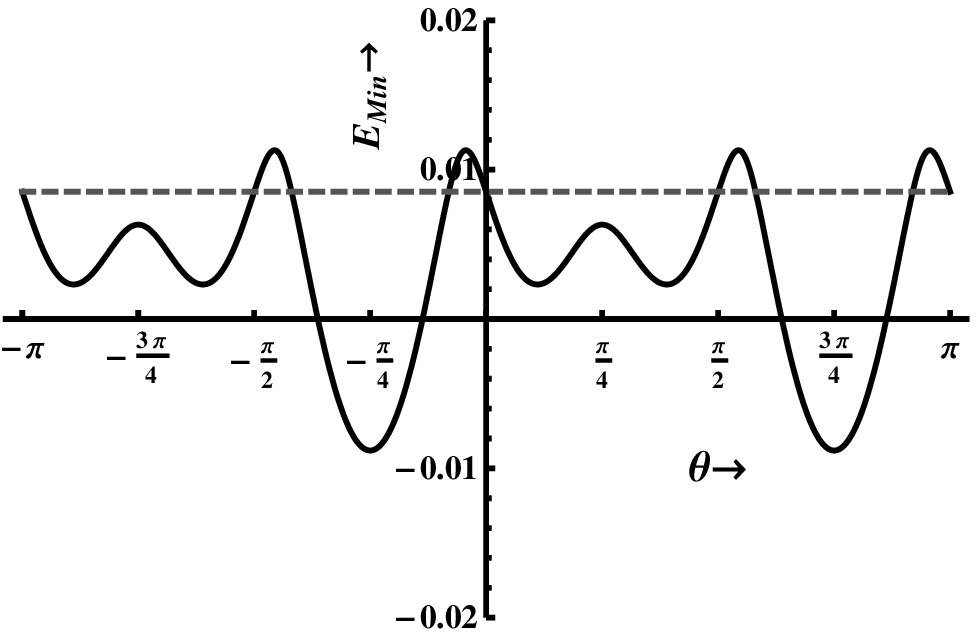}\qquad
\includegraphics[width=7cm]{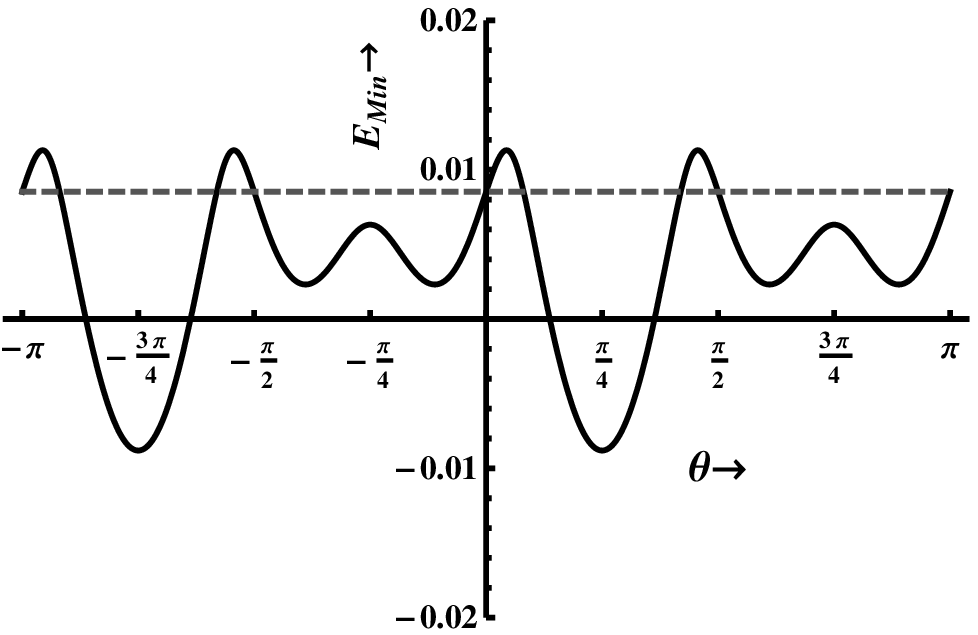}
\caption{
Plot of minimum eigenvalue as a function of $\theta$ of the
operators $\rho^{\prime}_{1}(\theta)$ and
$\rho^{\prime}_{2}(\theta)$ obtained after the action of
$I_3\otimes \varphi_{C_{1,2}}(\theta)$ on the state $\rho$
defined in equation~(\ref{upb_entang}). The upper graph
corresponds to $\rho^{\prime}_{1}(\theta)$ while the lower
one corresponds to $\rho^{\prime}_{2}(\theta)$.  The
straight line represents the minimum eigenvalue
corresponding to the operators obtained after the action of
the corresponding Choi map through the operators
$I_3\otimes\varphi_{C_{1,2}}$. The negativity of the minimum
eigen value, which occurs in both the graphs in a 
similar way but for shifted values of $\theta$, indicates that
the map has revealed the entanglement of the state. 
\label{upbplot1}
}
\end{figure}
Consider a one-parameter family of extremal
extensions of the Choi maps  
$\varphi_{C_{1,2}}(\theta)=\varphi_{C_{1,2}} \circ
U(\theta)$ with 
\begin{equation}
U(\theta)=\begin{pmatrix}
\cos\theta & 0 &\sin\theta \\
0 & 1 & 0\\
-\sin\theta & 1 & \cos\theta
\end{pmatrix}.
\label{unitary}
\end{equation}
These two families of maps defined via the unitary inner
automorphism can now be tried on the PPT entangled states
defined in Equation~(\ref{upb_entang}) to see if they can
reveal its entanglement.
We apply the maps $I\otimes \varphi_{C_{1,2}}(\theta)$ to the
state defined in Equation~(\ref{upb_entang}).
\begin{equation}
I_3\otimes \varphi_{C_{1,2}}(\theta)\, : \, \rho \rightarrow
\rho^{\prime}_{1,2}(\theta)  
\end{equation}
We compute the eigen values of $\rho^{\prime}_{1}(\theta)$
and $\rho^{\prime}_{2}(\theta)$. It turns out that 
the smallest eigen value becomes negative for a range of
$\theta$ values indicating that the resultant operator is
not a state, thereby revealing the entanglement of the
original state $\rho$.  
The plot of minimum eigen values of
$\rho^{\prime}_{1}(\theta)$ and $\rho^{\prime}_{2}(\theta)$
are shown in Figure~\ref{upbplot1}. The upper
graph corresponds to the case $\rho^{\prime}_{1}(\theta)$
and the lower one corresponds to the case
$\rho^{\prime}_{2}(\theta)$.

Both  families of maps are able to reveal the entanglement
of the state $\rho$ defined in Equation~(\ref{upb_entang}).
However the $\theta$ ranges  for which  the map reveals the
entanglement are different in each case.  The lower graph
can be superimposed on the upper graph by  a shift  of
$\frac{\displaystyle \pi}{\displaystyle 2}$ in $\theta$.  In
each graph the straight lines show the positive minimum
eigen value obtained after application of the corresponding
non-modified Choi map. 
\subsection{The PYRAMID construction}
Another interesting UPB construction 
for the $3\otimes 3$ Hilbert space is the PYRAMID 
construction~\cite{B2}.  We first define five vectors in
a three dimensional Hilbert space as:
\begin{equation}
v_i=N\left(\cos \frac{2\pi j}{5},\sin \frac{2\pi
i}{5},h\right) \quad j=0,\cdots,4; 
\end{equation}
where $h=\frac{1}{2}\sqrt{1+\sqrt{5}}$ and
$N=\frac{2}{\sqrt{5+\sqrt{5}}}$.
Using these vectors we define the UPB set as
\begin{equation}
|\psi_j\rangle=|v_j\rangle\otimes|v_{2j\mod
5}\rangle,\quad j=0,\cdots,4.
\label{pyramid}
\end{equation}
The corresponding PPT entangled state is obtained by 
substituting the UPB states given in 
Equation~(\ref{pyramid}) above into  Equation~(\ref{upb_entang}).
\begin{figure}[h]
{\includegraphics[width=7cm]{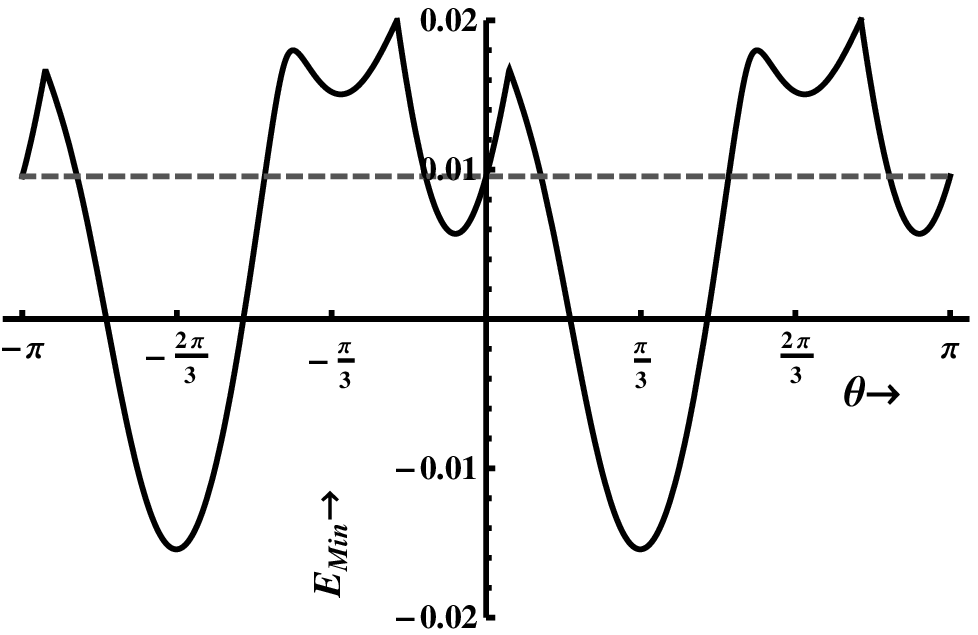}}\qquad
{\includegraphics[width=7cm]{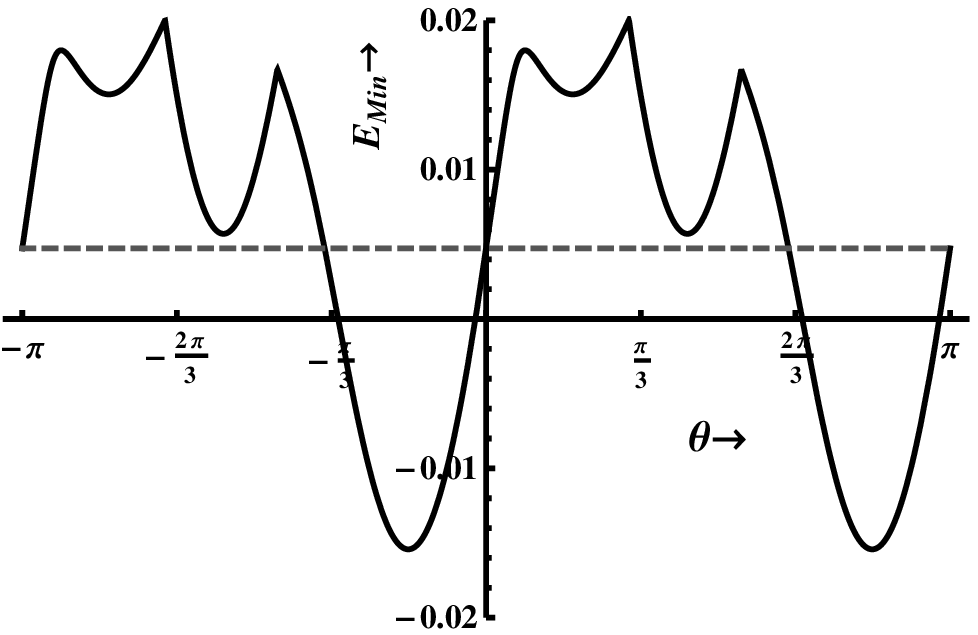}}
\caption{
Plot of minimum eigen value of operators 
$\rho^{\prime}_{1,2}(\theta)$ as a function of $\theta$.
The operators $\rho^{\prime}_{1,2}(\theta)$ are
obtained from the PPT entangled states in the orthogonal
complement of the PYRAMID UPB construction by the action of
families of extremal extensions of two Choi maps on the
second system. The negativity of the minimum eigenvalues
shows that the map is able to detect entanglement of the states.
The straight line in each graph shows the minimum eigenvalue 
in the case of the original Choi map which remains
positive and therefore does not reveal the entanglement.
}\label{upbplot2} 
\end{figure}
We carry out an identical analysis to the TILES case and
find that the entanglement of this state is again detected
by the modified Choi maps.  The plots are shown in
Figure~\ref{upbplot2} where the minimum eigen value is
displayed as a function of $\theta$ for the operator
obtained after action of modified Choi Both the  families of
maps reveal the entanglement of the state and the graphs
(Figure~\ref{upbplot2}) also display an invariance under a
shift of $\frac{\displaystyle \pi}{\displaystyle 2}$ in
$\theta$, as was seen for the TILES case.  However in this
case, the range of values over which the minimum eigen value
is negative is different. This means that the extremal maps
which reveal the entanglement of the state in this case are
different from the ones in the TILES case.  maps.

\section{Further examples of extermal extensions}
\label{examples}
To demonstrate the usefulness of the extensions based on
automorphisms we describe below three insightful results.
The first result is that the two maps due to Choi described
in Equations~(\ref{choi1}) and ~(\ref{choi2}) naturally get
connected via a
combination of inner and outer unitary automorphisms.  The
map  $\varphi_{C_1}$ thus gets related to  $\varphi_{C_2}$.  
\begin{equation}
\varphi_{C_1} = U\left(\frac{3\pi}{2}\right)
\circ \varphi_{C_2}\circ
U\left(\frac{\pi}{2}\right).
\end{equation}

Secondly, the construction that we had described in a recent
paper where we had generated extremal maps as candidate
entanglement witnesses from the existing ones turns out to
be a nonr-unitary inner automorphism~\cite{PhysRevA.84.032328}. 
To describe this connection 
we consider an  extremal positive in-decomposable map
$\varphi:\mathcal{B}(\mathbb{C}^n)\longrightarrow\mathcal{B}(\mathbb{C}^n)$
and the corresponding bi-quadratic  
$F\begin{pmatrix}
X\\Y\end{pmatrix}=F\begin{pmatrix}x_1 &\cdots &x_n\\y_1
&\cdots & y_n\end{pmatrix}=\langle Y|\varphi(|X\rangle\langle X|)
Y\rangle$, where $|X\rangle=(x_1,\cdots,x_n)^t$, and
$|Y\rangle =(y_1,\cdots,y_n)^t$, $t$ denotes the transpose,
and $x_i,y_j$ are real parameters. 
A map $\varphi$ is positive and extremal
if and only if the corresponding real bi-quadratic form is
positive and extermal. In-decomposability of the map
implies that the form $F$ can not be written as a sum of square
of quadratic forms. 
It was shown in~\cite{PhysRevA.84.032328} that for any set of $n$ non zero real
parameters $a_1,\cdots, a_n$; the
form $G\begin{pmatrix}x_1 &\cdots &x_n\\y_1
&\cdots & y_n\end{pmatrix}=F\begin{pmatrix}a_1x_1 &\cdots
&a_nx_n\\y_1
&\cdots & y_n\end{pmatrix}$ is also an extremal positive
form. Hence the corresponding map denoted by
$\varphi_{(a_1,\cdots,a_n)}$ is an extremal in-decomposable
positive map. We had used this extended class of maps to
unearth the entanglement of a new class of PPT entangled
states~\cite{PhysRevA.84.032328}.
It turns out that this extremal extension can
be recast as an inner automorphism of the original map given
below 
\begin{equation}
\varphi_{(a_1,\cdots,a_n)}=\varphi\circ A
\end{equation}
where $A$ is an operator given  by the diagonal matrix
\begin{equation}
A={\rm Diag}( 
a_1,a_2 \cdots, a_n)
\end{equation}
This is clearly a non-unitary inner automorphism and
connects our earlier result with the present formulation.

In the third example we turn to a generalization of the Choi map
defined by Cho et. al.~\cite{cho1} as
\begin{eqnarray}
&&\varphi_m((x_{ij}))\longmapsto\nonumber  
\frac{1}{2}\times\\
&&
\!\!\left[
\begin{array}{lcr}
ax_{11}+bx_{22}+cx_{33}\!& -x_{12} & -x_{13}\\
-x_{21} &\!\!\! \!ax_{22}+bx_{33}+cx_{11}\!\!\! & -x_{23}\\
-x_{31} & -x_{32} &\! \!\!\!ax_{33}+bx_{11}+cx_{33}\!
\end{array}
\right]
\nonumber\\
\label{kye}
\end{eqnarray}
where $a,b,c$ satisfy certain conditions given in
detail in their paper.

It has been shown by Ha and Kye~\cite{1108.0130} that a
sub-class of the above family of maps, given by 
\begin{equation}
0<a<1,\quad  a+b+c=2, \quad bc=(1-a)^2; 
\end{equation} 
are extremal maps.  
It has been further
shown~\cite{PhysRevA.84.024302,1108.0130}
that these extremal maps can be
written as a one-parameter family of maps $\varphi_t$ 
with  $0 \leq t < \infty$. The parameters $a(t), b(t)$ and
$c(t)$ are given by 
\begin{equation}
a(t)=\frac{(1-t)^2}{1-t+t^2},\,
b(t)=\frac{t^2}{1-t+t^2},\, c(t)=\frac{1}{1-t+t^2}.
\end{equation}
We have $\varphi_{t=0}=\varphi_{C_1}$,
$\varphi_{t\rightarrow\infty}=\varphi_{C_2}$ while 
$\varphi_{t=1}$ is a decomposable map.
Using the unitary automorphism defined through the
one-parameter family of unitary transformations
given in Equation~(\ref{unitary}), we are able to
relate the maps in the interval $[0,1]$ to maps
in the interval $[1,\infty)$ as follows:
\begin{equation}
\varphi_{t} = U\left(\frac{3\pi}{2}\right)\circ
\varphi_{\frac{1}{t}}\circ
U\left(\frac{\pi}{2}\right).
\end{equation}
This mean that we need to consider only the maps in the
interval $[0,1]$ if we are interested in using
them as entanglement witnesses and the others can
be generated via the automorphism given above.
The above examples show that the automorphisms provide
us with a way  to connect various seemingly unrelated
maps. 
\section{Concluding Remarks}
\label{conclusions}
In this paper we have described extremal extensions of P
maps which are not CP via their composition with quantum
operations. Two  kinds of automorphisms are described and it
is shown that only one of them, namely, the inner
automorphism has the ability to enhance the entanglement
detection power of the original map. This construction opens
up new  possibilities of extremal extensions of P maps which
are CP. Focusing on the famous Choi map and its extensions
via a one-parameter family of unitary transformations, we
have discovered a useful and interesting connection with UPB. We
discover that for a certain parameter range the map begins
to unearth the entanglement of states in the orthogonal
complement of UPB.

The exposedness of maps has been discussed and
used in the entanglement context in a recent
interesting development~\cite{1108.0130}. It turns
out that the automorphisms described in our work
preserve the exposed property and thus if we start
with an exposed map we can construct families of
exposed maps. In this context extensions of
extermal exposed maps have also been considered by
Sarbicki and Chru{\'s}ci{\'n}ski~\cite{chur}.

In the context of UPB there is a way to interpolate between
TILES and PYRAMID~\cite{T2}. This possibility  provides us
with a rich variety of PPT entangled states. The
possibility of detecting these states with extensions of
already known P but not CP maps or implicating the non-CP
character of certain maps using these states will be taken
up elsewhere.  There could be interesting consequences of
these results in higher dimensions and they will also be
taken up elsewhere.  

%
\end{document}